\documentclass[10pt,twocolumn,a4paper]{article}

\usepackage[left=2cm,right=2cm,top=3cm,bottom=4cm,includeheadfoot]{geometry}
\usepackage{color,colortbl}
\usepackage{nicefrac}
\usepackage{listings}
\usepackage{amsfonts}
\usepackage{amsmath}
\usepackage{amssymb}
\usepackage{amsthm}
\usepackage{mathtools}
\usepackage[dvips,pdftex]{graphicx}
\usepackage{cite}
\usepackage{nicefrac}
\usepackage{authblk}
\usepackage{abstract}
\usepackage{siunitx}
\usepackage[pdfpagelabels,pdftex]{hyperref}
\usepackage{multirow} 
\usepackage{booktabs}

\usepackage[latin9]{inputenc}
\title{Isotope shift of $^{40,42,44,48}$Ca in the $4s\,^2\mathrm{S}_{\nicefrac{1}{2}} \rightarrow 4p\,^2\mathrm{P}_{\nicefrac{3}{2}}$ transition}

\author[1]{C.~Gorges}
\author[2]{K.~Blaum}
\author[3]{N.~Fr\"ommgen}
\author[3]{Ch.~Geppert}
\author[3]{M.~Hammen}
\author[1]{S.~Kaufmann}
\author[1]{J.~Kr\"amer}
\author[1]{A.~Krieger}
\author[2]{R.~Neugart}
\author[4]{R.~Sanchez}
\author[1,3]{W.~N\"ortersh\"auser}
\date{}
\affil[1]{Institut f\"ur Kernphysik, Technische Universit\"at Darmstadt, Germany}
\affil[2]{Max-Planck-Institut f\"ur Kernphysik, Heidelberg, Germany}
\affil[3]{Institut f\"ur Kernchemie, Johannes Gutenberg-Universit\"at Mainz, Germany}
\affil[4]{GSI Helmholtzzentrum f\"ur Schwerionenforschung, Darmstadt, Germany}

\begin{document}
\twocolumn[{%
  \maketitle
\begin{abstract}
We report on improved isotope shift measurements of the isotopes $^{40,42,44,48}$Ca in the $4s\; ^2\mathrm{S}_{\nicefrac{1}{2}} \rightarrow 4p\, ^2\mathrm{P}_{\nicefrac{3}{2}}$ (D2) transition using collinear laser spectroscopy. Accurately known isotope shifts in the \nolinebreak{$4s\; ^2\mathrm{S}_{\nicefrac{1}{2}} \rightarrow 4p\; ^2\mathrm{P}_{\nicefrac{1}{2}}$} (D1) transition were used to calibrate the ion beam energy with an uncertainty of $\Delta U \approx \pm 0.25$\,V. The accuracy in the D2 transition was improved by a factor of $5 - 10$. A King-plot analysis of the two transitions revealed that the field shift factor in the D2 line is about 1.8(13)\,\% larger than in the D1 transition which is ascribed to relativistic contributions of the $4p_{1/2}$ wave function.  
\end{abstract}
  \bigskip
}]
\section{Introduction}
Hyperfine structure and isotope shifts have been studied in various transitions of calcium  isotopes for decades both experimentally (see e.g. \cite{Silverans, Palmer, Vermeeren, Mortensen, Brun, Lucas, Bini, Salum,  Grundevik, Wolf2009, Andl, Gade}) and theoretically (e.g. \cite{Caurier, Holt, Roth, Soma, Hagen, Holt2}). This is due to the importance of this element in various fields: With $^{40}$Ca and $^{48}$Ca the calcium isotopic chain is the only one that contains two naturally occurring doubly magic isotopes. Moreover, the nuclear charge radii along these stable isotopes show a very peculiar behaviour, having a maximum at $^{44}$Ca, a strong odd-even staggering and almost identical charge radii for $^{40}$Ca and $^{48}$Ca \cite{Wohlfahrt, Emrich} despite the large difference in nuclear mass. The isotopic chain has therefore been of great interest for nuclear structure physics. 
Additionally, the long-lived isotope $^{41}$Ca can be used for dating applications \cite{Brown} as well as a tracer in biology and medicine to study the physiology of calcium in the human body \cite{Denk, Turteltaub, Vogel}. Hence, isotope shifts and hyperfine structure were also studied in several atomic transitions \cite{Noer98, Mueller00} for isotope selective resonance ionization and ultra-trace analysis \cite{Lu}. The calcium ion is a workhorse in the field of quantum-optical applications in ion traps and has thus also been studied with high accuracy\cite{Wolf2009, Wan, Gebert}. Moreover, the isotope shift information in the $3d \; ^2D_J \rightarrow 4p \; ^2P_J$ infrared triplet \cite{Noertershaeuser} has led to the discovery of an anomalous isotopic composition in mercury-manganese (HgMn) stars, in which the isotopic Ca ratio in the stellar atmosphere is dominated by $^{48}$Ca \cite{Castelli2004,Cowley2005}.\\
Precision data are available for many transitions, but the isotope shift in the $4s\; ^2\mathrm{S}_{\nicefrac{1}{2}} \rightarrow 4p\, ^2\mathrm{P}_{\nicefrac{3}{2}}$ transition of calcium, often called D2 line\footnote{since it is isoelectronic to the corresponding Fraunhofer D2 line in Na}, is only known with moderate accuracy \cite{Mart}, even though this transition is preferred for studies of nuclear moments in short-lived isotopes. It allows for an extraction of the nuclear spectroscopic quadrupole moment from the electric hyperfine structure, in contrast to the $4s\; ^2\mathrm{S}_{\nicefrac{1}{2}} \rightarrow 4p\, ^2\mathrm{P}_{\nicefrac{1}{2}}$ (D1) transition. This is of special importance because of the revived interest of nuclear structure physics in spectroscopic information on stable and short-lived calcium isotopes. The interest concentrates on the neutron-rich region where new shell closures have been observed recently for the isotopes $^{52,54}$Ca at neutron numbers $N=32$\cite{Wienholtz} and $N=34$ \cite{nature}. These shell closures are associated with three-nucleon forces and can be studied theoretically using chiral effective field theory \cite{Hagen, Holt3}. In order to provide additional information about these shell closures, nuclear moments and isotope shifts of neutron-rich calcium isotopes have been studied at ISOLDE/CERN. The COLLAPS collaboration has measured hyperfine structure and isotope shifts beyond $N=28$, namely the isotopes $^{49-52}$Ca. Results for the nuclear moments have already been reported \cite{ronald}, while the analysis of the isotope shifts is hampered by the large uncertainties  in the D2 isotope shifts of the stable isotopes. These data are required because of the large measuring uncertainty of the exact high voltage potential of the RFQ cooler and buncher ISCOOL \cite{Franberg, ManeISCOOL}. This device is used to collect  the ions delivered from the ISOLDE on-line ion source and to send them in short bunches towards the collinear laser spectroscopy setup. 
The ion beam energy is usually the limiting systematic uncertainty for isotope shift measurements in collinear laser spectroscopy. But if accurate isotope shifts are available for several isotopes, the beam energy can be calibrated using this information.
In order to provide the required reference data for the analysis of the short-lived isotopes, we have performed  isotope shift measurements of $^{40,42,44,48}$Ca$^+$ at the TRIGA-LASER setup in the Institute of Nuclear Chemistry in Mainz \cite{ket}.
High-accuracy data of the isotope shifts in the $4s_{\nicefrac{1}{2}} \rightarrow 4p_{\nicefrac{1}{2}}$ transition, recently obtained using logical spectroscopy in an ion trap with an accuracy better than $100$\,kHz \cite{Gebert}, were used for a reliable calibration of the ion source potential. In this way we have extracted the isotope shifts in the D2 line with up to tenfold improved accuracy. In addition, the two transitions D1 and D2 can be compared via a King plot to examine the effect of the electron density of the $p_{\nicefrac{1}{2}}$ electron at the nucleus. Similar studies have previously been performed for instance in Ba \cite{Wendt}, Mg \cite{Batteiger} and Ra \cite{Neu}.
\section{Experimental setup}
The TRIGA-LASER experiment is part of the TRIGA-SPEC setup at the TRIGA research reactor Mainz \cite{ket}. The reactor will provide exotic nuclei produced by neutron-induced fission inside the reactor core for 'online' collinear laser spectroscopy measurements.
\begin{figure}[h]
	\centering
		\includegraphics[width=0.49\textwidth]{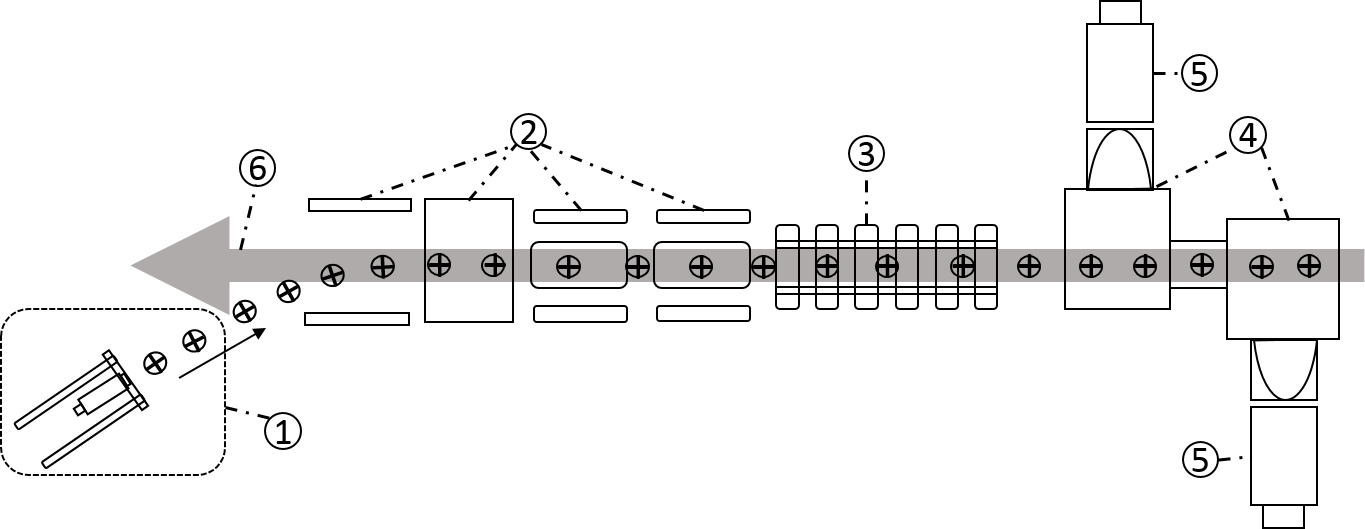}
	\caption{Schematic drawing of the main parts of the TRIGA-LASER beamline. 1. "high voltage cage" with offline ion source; 2. ion steering elements; 3. charge exchange cell (not used); 4. mirror system for fluorescence detection; 5. Photomultiplier tubes; 6. laser beam}
	\label{fig:TRILA}
\end{figure}

The measurements on stable calcium isotopes presented in this work were performed on ion beams of typically $0.8$~nA with the natural isotopic composition produced in an offline ion source: A graphite tube with a central bore was filled with a few grains of metallic calcium. This crucible was heated by applying a current of about 50\,A causing a voltage drop of about $4$~V along the crucible. The complete ion source is mounted on a high-voltage platform to which a potential of about $10~\rm{kV}$ is applied. The ions are accelerated towards ground potential. The ion source voltage is generated by a Heinzinger PNChp60000-10ump power supply with a relative stability specified to be better than $10^{-5}$. The potential of the high-voltage platform is continuously monitored and recorded using a Julie Research KV-10R high-voltage divider.
The divider ratio is known with a relative precision of $5 \times 10^{-5}$ and a high-precision voltmeter of the type Agilent 34401A is used to measure the voltage across the precision resistor at the end of the divider chain. The divider ratio has been determined by comparison with the KATRIN high-voltage divider \cite{Thuemmler,Bauer} at the University of M\"unster. The starting potential of the ions inside the graphite furnace can differ from the platform potential by up to 4\,V due to the voltage drop caused by direct heating of the graphite furnace.
A $10^\circ$ deflector is used to superimpose the uv-laser beam with the ion beam in anticollinear geometry.
The beamline is equipped with two pairs of deflector electrodes in vertical and horizontal direction and a quadrupole doublet for beam steering and shaping.
Superposition of the two beams is ensured using several apertures with different sizes at two positions of the beamline.\\
The optical detection region is supplied with two mirror systems for light collection and appropriate photomultiplier tubes. They are mounted on an insulated platform to allow for Doppler-tuning by accelerating or decelerating the ions into the detection region. An offset potential of up to $\pm 10$\,kV can be applied to the platform by a high-voltage device with similar  specifications as described above. This supply is again mounted on a platform and can be raised in potential by a Kepco BOP~500M fast high-voltage amplifier. This allows fast Doppler scanning across $\pm 500$\,V, corresponding to almost 30\,GHz scanning range in the case of calcium and 10\,kV beam energy. The voltages applied to this platform during the measurements were also continuously monitored and recorded by a high-voltage divider and an Agilent 34401A.\\
A Matisse TS titanium:sapphire (Ti:Sa) ring-laser pumped by a Verdi Nd:YVO$_4$ laser is used to produce infrared light at $794~\rm{nm}$ and $786~\rm{nm}$ for excitation of the $4s_{\nicefrac{1}{2}} \rightarrow 4p_{\nicefrac{1}{2}}$ and the $4s_{\nicefrac{1}{2}} \rightarrow 4p_{\nicefrac{3}{2}}$ transition, respectively, after second-harmonic generation in a Wavetrain frequency doubling cavity. The Ti:Sa laser is actively stabilized to an external cavity to correct for short-term fluctuations and to a FC1500 frequency comb to ensure long-term stability and an accurately known laser frequency during the measurements. The linewidth of the laser is approximately 250\,kHz per 4\,hours. The uv light is coupled to a 180\,m long single-mode fiber and transported to the beamline. The beam leaving the fiber is collimated, beam-halo is removed  by a spatial mode filter to reduce scattered light background and coupled into the beamline through a quartz window. The light leaves the vacuum system through a second window on the opposite side of the beamline. The light power was kept below 150~$\mu$W in a beam of approximately 1.5\,mm diameter and no appreciable sign of saturation broadening was observed at these intensities. With the optical detection setup composed of two mirror systems based on non-imaging optics an efficiency of one detected fluorescence photon per 200 calcium ions has been demonstrated for the even isotopes without hyperfine splitting.
\section{Experimental procedure}
We have performed measurements in the $4s_{\nicefrac{1}{2}} \rightarrow 4p_{\nicefrac{1}{2}}$ transition before and after the measurements in the $4s_{\nicefrac{1}{2}} \rightarrow 4p_{\nicefrac{3}{2}}$ line. All measurements were performed at a constant starting potential without changing the ion optical settings and with the same laser system.
For switching between the two transitions, the laser was changed in wavelength and the SHG cavity was optimized again. The fiber transport of the laser beam ensured that the alignment of the laser through the beamline did not change.
The laser frequencies for both transitions were chosen such that comparable voltages had to be applied to the optical detection region for each isotope in order to avoid systematic changes of beam focussing.
To increase the measurement accuracy, spectra of the heavier calcium isotopes were always alternated with reference measurements of $^{40}$Ca$^+$.
\begin{figure}[h]
	\centering
		\includegraphics[width=0.49\textwidth]{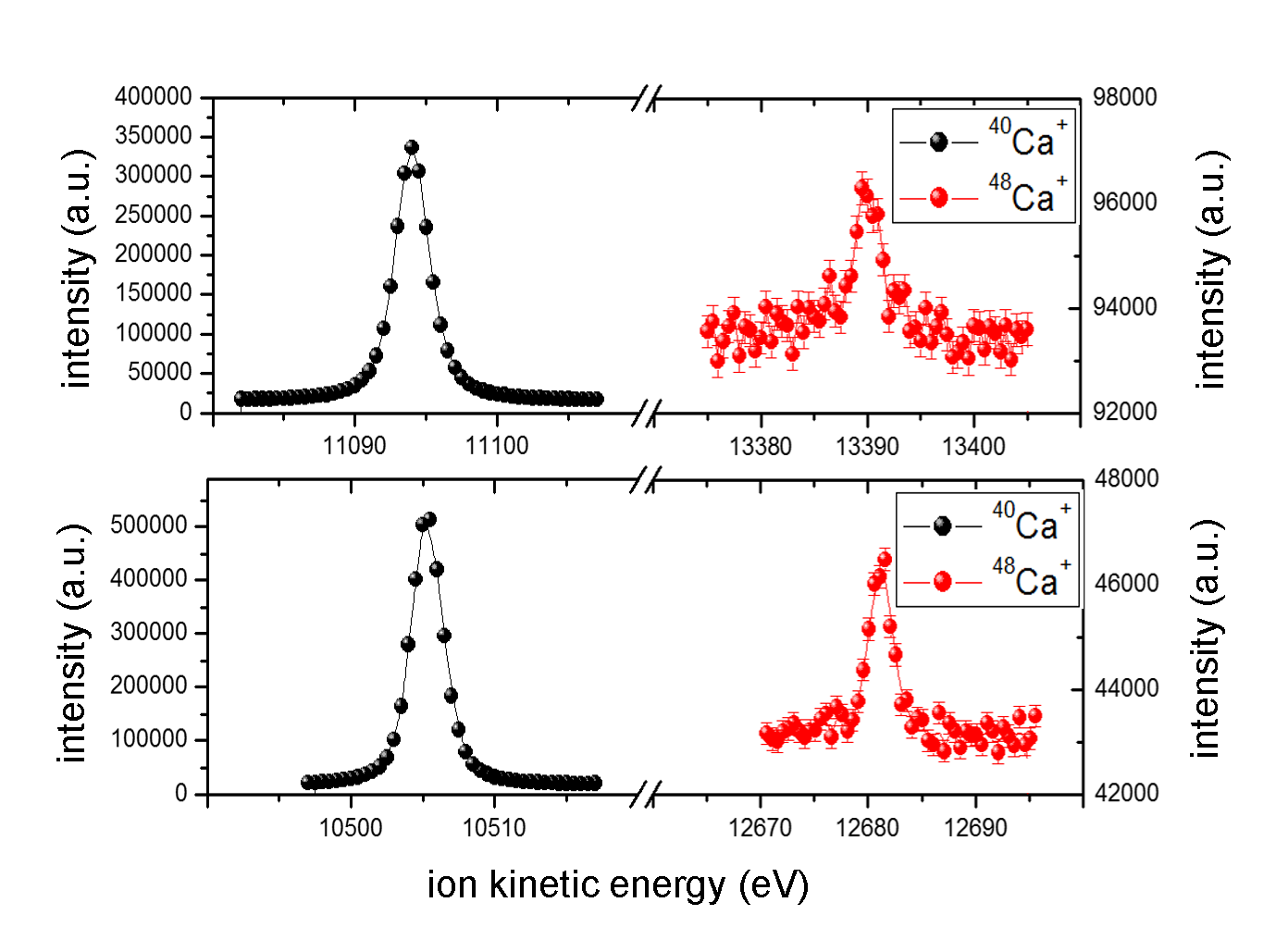}
	\caption{Typical spectra of $^{40}$Ca$^+$ and $^{48}$Ca$^+$. The upper panel shows the $4s_{\nicefrac{1}{2}} \rightarrow 4p_{\nicefrac{3}{2}} $ transition (D2) and the lower one the $4s_{\nicefrac{1}{2}} \rightarrow 4p_{\nicefrac{1}{2}}$ transition (D1).}
	\label{fig:lineshapes}
\end{figure}
\section{Analysis}
Line profiles of $^{40}$Ca$^+$ and $^{48}$Ca$^+$ in both transitions are depicted in Fig.\,\ref{fig:lineshapes}. In the analysis of the spectra, for each data point the total acceleration voltage was converted into the corresponding Doppler-shifted frequency in the rest frame of the Ca$^+$ ion. All lines were then fitted with Voigt profiles using a fixed Lorentzian width of $\Delta\nu_\mathrm{L} = 23.4\,\rm{MHz}$ (natural linewidth). Gaussian linewidth $\Delta\nu_\mathrm{G}$, amplitude, background and peak center position $\nu_0$ were used as free parameters. The linewidth parameter $\Delta\nu_\mathrm{G} \approx 17 - 20$\,MHz was obtained from the fitting routine, with little changes between the isotopes due to the different masses.\\
A slight asymmetry that is observed in the peak shape was similar for all isotopes. Tests with different lineshapes that better fitted to the profile, did not result in any significant change in the isotope shift.
\subsection{Beam energy calibration}
\begin{figure}[h!]
	\centering
		\includegraphics[width=0.49\textwidth]{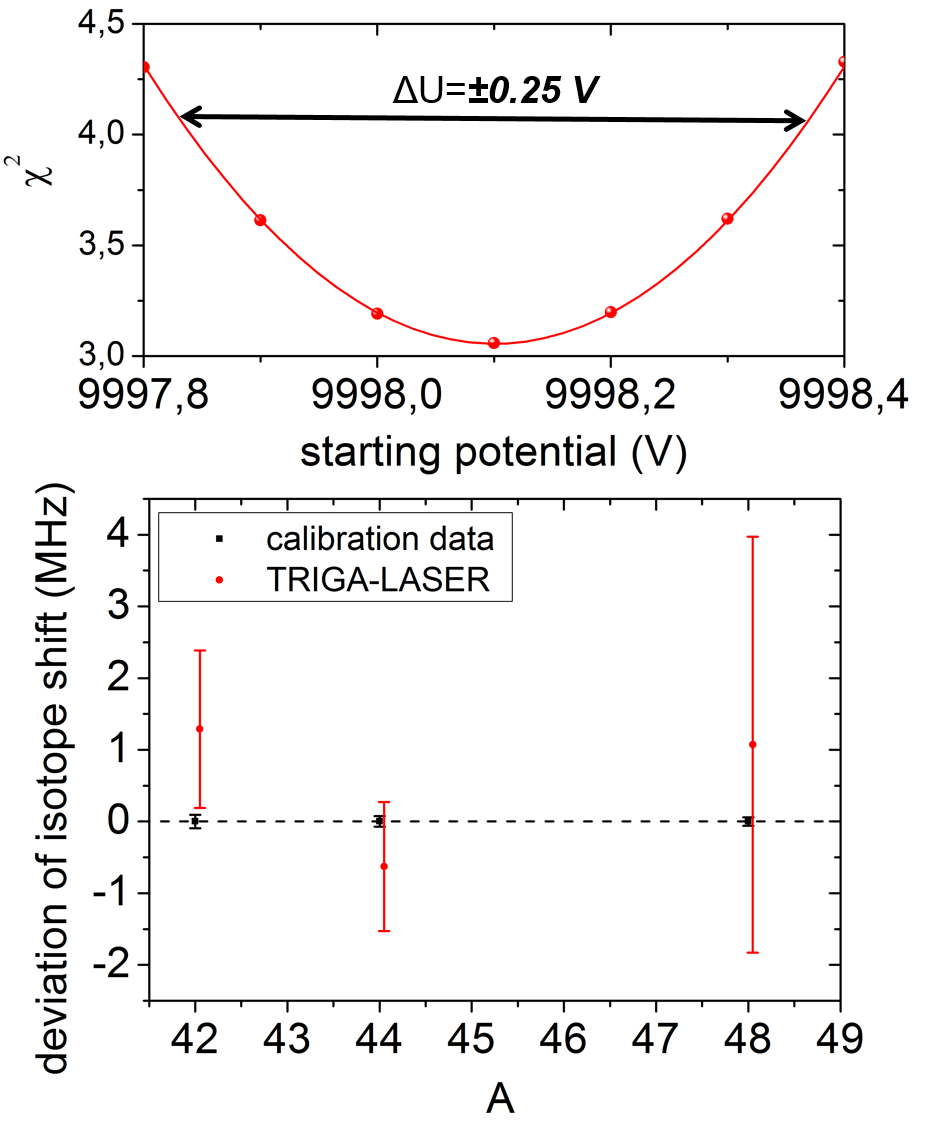}
	\caption{In the upper panel the $\chi^2$-values according to Eq.\,\ref{eq:chi} are plotted for different assumptions of the acceleration voltage. The deviation between the obtained isotope shifts in the D1 transition and the calibration data \cite{Gebert} (black squares, set to zero) are plotted for the different isotopes in the lower panel. Error bars are purely statistical uncertainties.}
	\label{fig:D1conclusion}
\end{figure}

To increase the measurement accuracy, resonance scans of the isotope of interest $^X$Ca$^+$ ($X=42,44,48$) and the reference isotope $^{40}$Ca$^+$ were alternated such that two scans of $^X$Ca$^+$ were interleaved by four scans of $^{40}$Ca$^+$ to reduce the influence of long-term fluctuations. The whole measuring cycle (cycle I) was repeated at a day about three months later (cycle II). In the meantime the beamline was modified at different points and everything was readjusted.
As described above, the starting potential is known to $U_{ \mathrm{source}} \, _{-1 \, \rm{V}}^{+ 4 \, \rm{V}}$, where the uncertainties are determined by the accuracy of the voltage measurement ($\Delta U/U \approx 10^{-4}$) and the voltage drop along the crucible that will always lead to a higher voltage than that applied to the platform. To reduce the uncertainty, the resulting isotope shifts $\delta \nu_{\mathrm{IS}}^{A,40} (U_{\mathrm{start}})$ of all isotopes in the D1 line for various starting potentials $U_{\mathrm{start}}$ between $9996$~V and $10$~kV were derived and for each voltage the mean square deviation from the precise calibration data $\delta \nu_{\mathrm{trap}}^{A,40}$ 
\begin{equation}
\chi^2 (U_{\mathrm{start}}) = \sum\limits_{A=42,44,48} \left( \frac{\delta \nu_{\mathrm{IS}}^{A,40} (U_{\mathrm{start}}) - \delta \nu_{\mathrm{trap}}^{A,40}}{\Delta\delta \nu_{\mathrm{IS}}^{A,40}} \right)^2 \, ,
\label{eq:chi}
\end{equation}
with the purely statistical uncertainty $\Delta\delta \nu_{\mathrm{IS}}^{A,40}$, was calculated.
The value of $\chi^2$ obtained as a function of the starting potential is plotted in Fig.~\ref{fig:D1conclusion} for the case of the second cycle. The $\chi^2$ data are very well fitted by a parabola and the minimum of the curve is the most probable starting potential. The uncertainty is estimated to be $0.25$~V, which corresponds to an increase of $\chi^2$ by 1 as indicated in Fig.~\ref{fig:D1conclusion}. The results are $9997.85(25) \, \mathrm{V}$ for the first and $9998.10(25) \, \mathrm{V}$ for the second measuring cycle. The small change in the starting potential is within the uncertainty of the results and might actually be due to a shifted ionization region for the ions within the graphite furnace caused by, e.g., different filling levels with calcium.\\
The results of the calibration measurements are presented in the lower panel of Fig.\,\ref{fig:D1conclusion}.
Shown are the statistical uncertainties. The absolute values of the isotope shifts can be found in Table~\ref{tab:IsotopeShifts}.
\subsection{Results}
With the calibrated starting potential we can derive the isotope shifts in the $4s_{\nicefrac{1}{2}} \rightarrow 4p_{\nicefrac{3}{2}}$ transition. All measurements of $^{40,42,44,48}$Ca$^+$ of both cycles have been combined to yield the isotope shift values listed in Table~\ref{tab:IsotopeShifts}.
The first bracket for each value shows the statistical uncertainty which is the standard deviation of all measurements and the second bracket represents the systematic uncertainty. The latter depends mainly on the two voltage uncertainties: the one of the starting potential $ \Delta U_{\mathrm{start}}$ and the one of the voltage floating the optical detection region $ \Delta U_{\mathrm{od}}$. The systematic uncertainty is derived as \cite{Mueller}:\\
\begin{equation}
\Delta \delta \nu_{\mathrm{IS}}^{\rm{A, A'}} = \nu_L \sqrt{\frac{e U_{\mathrm{start}}}{2m c^2}} \cdot \left( \Delta U  + \Delta M \right) \, ,
\end{equation}
with 
\begin{equation}
\Delta U = \frac{1}{2} \left( \frac{U_{\mathrm{od}}}{ U_{\mathrm{start}}} + \frac{\delta m}{m} \right) \cdot \frac{\Delta U_{\mathrm{start}}}{U_{\mathrm{start}}} + \frac{U_{\mathrm{od}}}{U_{\mathrm{start}}} \frac{\Delta U_{\mathrm{od}}}{U_{\mathrm{od}}}
\end{equation}
and
\begin{equation}
\Delta M  = \frac{\Delta m_{\rm{A}}}{m} +  \frac{\Delta m_{\rm{A'}}}{m}
\end{equation}
and the reference mass $m$, the mass difference $\delta m = m_{\rm{A'}} - m_{\rm{A}}$, the speed of light $c$ and the laser frequency  $\nu_L$. The isotope masses $m_{\rm{A, A'}}$ have been derived from the AME2012 atomic-mass evaluation table \cite{ame}, taking into account the mass of the missing electron while the missing binding energy of the electron needs not to be considered here.
\begin{figure}[h!]
	\centering
		\includegraphics[width=0.5\textwidth]{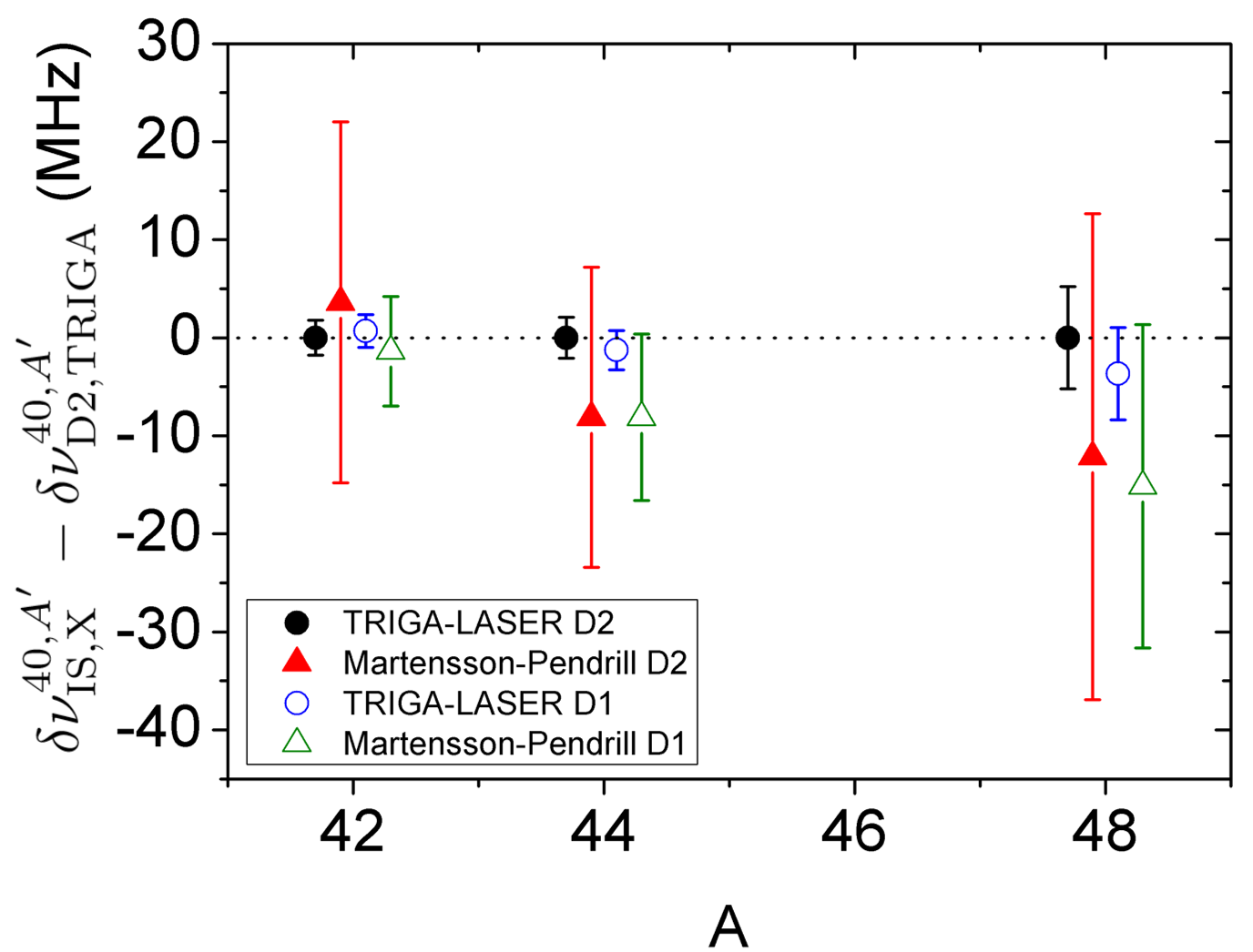}
	\caption{
	Deviations of D1 and D2 isotope shifts measured in this work and in Martensson-Pendrill et al. \cite{Mart} from the present values $\delta \nu^{40,A'}_{\rm{D2, TRIGA}}$.}
	\label{fig:D2conclusion}
\end{figure}

The offset voltages were measured continuously using a voltage divider providing a relative uncertainty of the respective acceleration voltage of $\Delta U_{\mathrm{od}}$/$U_{\mathrm{od}} = 5 \times 10^{-5}$.\\
The determined isotope shifts are shown in Fig.~\ref{fig:D2conclusion} where the error bars represent the total uncertainty which was obtained as the geometrical average of the statistical and the systematic uncertainty. The agreement with literature as well as the significant improvement in accuracy is clearly visible.
\begin{table}[h!]
	\centering
	\caption{Isotope shift measurements in two transitions of singly ionized calcium with respect to $^{40}\rm{Ca}^+$. The first bracket gives the statistical, the second one the systematic uncertainty.} 
		\begin{tabular}{c|l|l|l}
		\hline
		\hline
		A & {$\delta \nu_{4s_{\nicefrac{1}{2}} \rightarrow 4p_{\nicefrac{1}{2}}}$} & {$\delta \nu_{4s_{\nicefrac{1}{2}} \rightarrow 4p_{\nicefrac{3}{2}}}$} & Ref\\
		 & {(MHz)} & {(MHz)} & \\
		\hline
		$40$ & {$0$} & {$0$} & \\
		\hline		
		42 & 427.0 (11)$(13)$ & 426.4(15)$(10)$ & \\
		& 425(4)$(4)$ & 430 (18)$(4)$ & \cite{Mart} \\
		& 425.706(94) & &  \cite{Gebert} \\
		\hline
		$44$ & 848.8(9)$(18)$ &  850.1(10)$(19)$ & \\
		& 842(3)$(8) $ &  842(13)$(8)$ & \cite{Mart}\\
		& 849.534(74) & & \cite{Gebert} \\
		\hline
		$48$ & 1707.2 (29)$(38)$ &  1710.6(35)$(39)$ & \\
		& 1696(4)$(16)$ & 1699(19)$(16)$ & \cite{Mart}\\
		& 1705.389(60) & & \cite{Gebert} \\
		\hline
		\hline
		\end{tabular}
	\label{tab:IsotopeShifts}
\end{table}

\begin{figure}[h]
	\centering
		\includegraphics[width=0.5\textwidth]{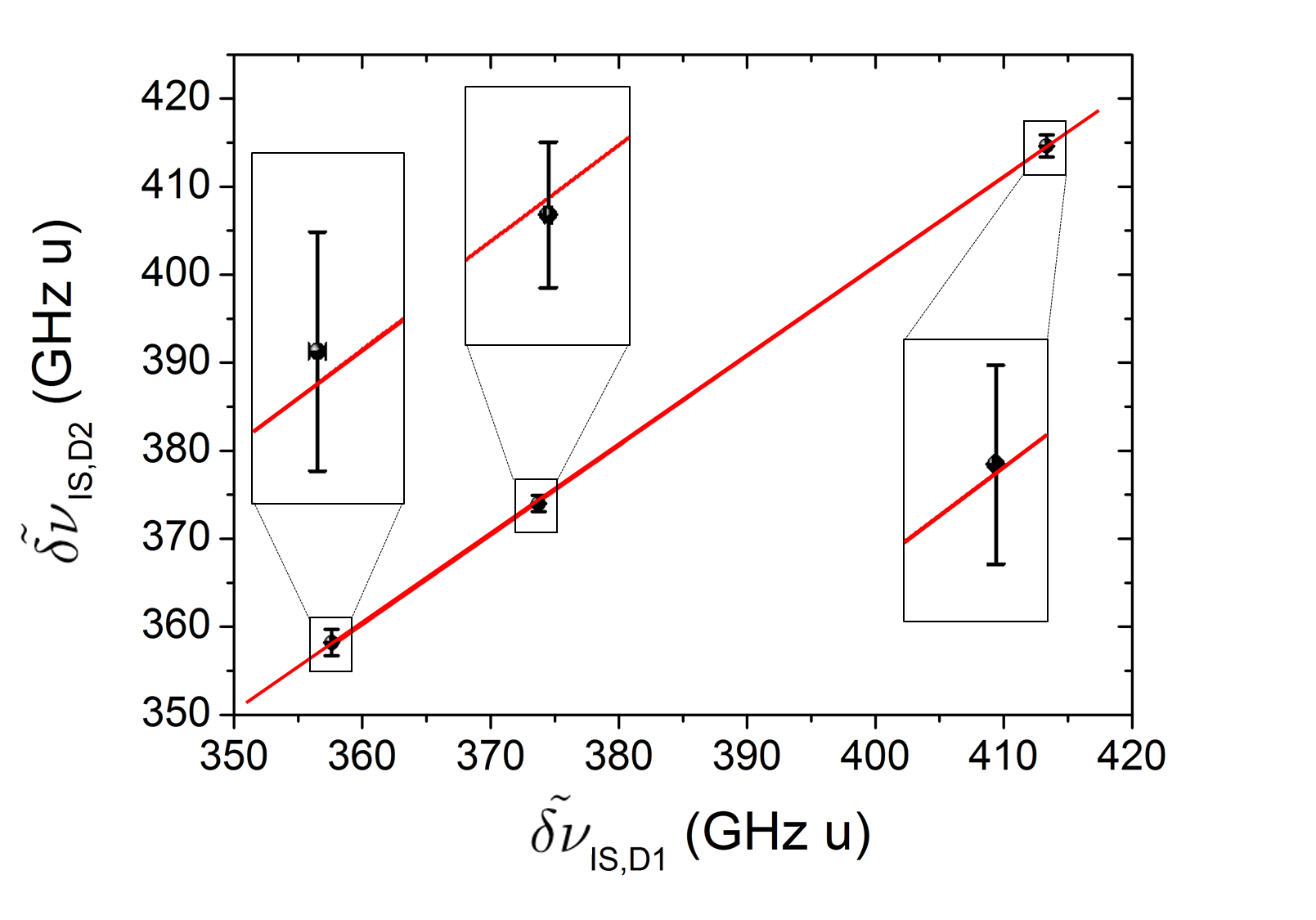}
	\caption{King plot showing the linear relation of the modified isotope shifts $\tilde{\delta \nu}_{\rm{IS}}$ in the $4s_{\nicefrac{1}{2}} \rightarrow 4p_{\nicefrac{3}{2}}$ (D2) and the $4s_{\nicefrac{1}{2}} \rightarrow 4p_{\nicefrac{1}{2}}$ (D1) transitions. The insets enlarge the data points to show the agreement with the fitted line.}
	\label{fig:kingplot}
\end{figure}

The isotope shift $\delta \nu^{A,A'}_{\rm{IS}}$ between two isotopes $A$ and $A'$ is caused by the change in mass and volume of the nuclei and can be split into the mass shift (MS) and the field shift (FS). 
\begin{equation}
\delta \nu^{A,A'}_{\rm{IS}} = \delta \nu^{A,A'}_{\rm{MS}} + \delta \nu^{A,A'}_{\rm{FS}} \ .
\end{equation}
The functional relation
\begin{equation}
\delta \nu^{A,A'}_{\rm{IS}} = F \cdot \delta \left< r^2 \right>^{A,A'} + M \cdot \frac{m_{A'} - m_{A}}{m_{A}\cdot m_{A'}} \ ,
\label{eq:IS}
\end{equation}
with field shift factor $F$, mass shift constant $M$ and change in mean square charge radius $\delta \left< r^2 \right>^{A,A'}$ is well known (see e.g. \cite{Mart}).
Both, the field and the mass shift depend on the investigated transition, but $\delta \left< r^2 \right>^{A,A'}$ is identical in both transitions. Hence it can be eliminated by comparing two transitions $i$ and $j$. Eq.~(\ref{eq:IS}) can thus be transformed into a linear relation between the isotope shifts in two transitions\cite{King, Heilig}:
\begin{equation}
\tilde{\delta \nu}^{A,A'}_{i} = \tilde{\delta \nu}^{A,A'}_{j} \frac{F_{i}}{F_{j}} + \left( M_{i} - M_{j} \cdot \frac{F_{i}}{F_{j}} \right) , 
\label{eq:king}
\end{equation}
where
\begin{equation}
\tilde{\delta \nu}^{A,A'}_{i/j}  = \frac{m_{A} \cdot m_{A'}}{m_{A}-m_{A'}} \delta \nu^{A,A'}_{i/j}
\end{equation}
is the modified isotope shift.\\
The modified isotope shifts in the D2 line determined here are plotted in Fig.~\ref{fig:kingplot} as a function of the modified isotope shift in the D1 line as measured in the ion trap~\cite{Gebert}. The data show an excellent linear behaviour as expected from Eq.~\ref{eq:king}. From the linear regression we obtained the ratio of the field shift factors of $F_{D2}$/$F_{D1} =1.018(13)$ which deviates from unity by $1.5~\sigma$.
The field shift factor $F$ is proportional to the difference of the electron probability density at the nucleus
\begin{equation}
\Delta \left| \Psi \right|^2 = \left| \Psi_{4s_{\nicefrac{1}{2}}} \right|^2 - \left| \Psi_{4p_{\nicefrac{1}{2},\nicefrac{3}{2}}} \right|^2
\end{equation}
and might therefore be slightly smaller for D1 since the $p_{\nicefrac{1}{2}}$ electron can obtain a small admixture of $s_{\nicefrac{1}{2}}$ to the wavefunction. This would result in a ratio slightly larger than 1 as observed. For a firm statement, a higher precision is needed.

From the $y$-axis intersection we can obtain the "weighted" difference between the specific mass shifts, which is
\begin{equation}
M_{D2} - M_{D1} \cdot \frac{F_{D2}}{F_{D1}} = -6.4(51)\, \mathrm{GHz \, u}
\end{equation}
as well compatible with zero within $1.3 \sigma$.
In order to improve the accuracy of the field shift factor ratio and the mass shifts, more precise measurements in a Paul trap will be carried out in the D2 line \cite{Piet}.
\section{Conclusion}
We have performed precision measurements of the isotope shifts in the D1 and the D2 transition in singly ionized calcium. The results agree with existing literature values and improve the accuracy in the D2 line considerably. They provide important reference data for isotope shift values of stable calcium ions and helpful calibration data for collinear laser spectroscopy measurements on the neutron-rich isotopes beyond $^{48}$Ca investigating the $4s \; ^2P_{\nicefrac{1}{2}} \rightarrow 4p \; ^2P_{\nicefrac{3}{2}}$ transition.\\
The field shift in the D2 line is 1.8(13)\,\% larger than in the D1 line. This observation might be used as a benchmark for relativistic atomic structure calculations. 
\section{Acknowledgments}
We acknowledge financial support from the German Federal Ministry for Education and Research (BMBF)  under contract 05P12UMFN8, the Helmholtz Association of National Research Centers  under contract VH-NG-148, the Helmholtz International Center for FAIR (HIC for FAIR) within the LOEWE program by the State of Hesse, and the Precision Physics, Fundamental Interactions and Structure of Matter Cluster of Excellence (PRISMA).  N.F. received support through GRK Symmetry Breaking (DFG/GRK 1581),  M.~H. and S.~K. from HGS-Hire and A.~K. from the Carl-Zeiss-Stiftung (AZ:21-0563-2.8/197/1).

\end{document}